**Electron phonon coupling in ultrathin Pb films on Si(111): Where the heck is the energy?**

M. Tajik[1], T. Witte[1], Ch. Brand[1], L. Rettig[3], B. Sothmann[1,2], U. Bovensiepen[1,2], M. Horn von Hoegen[1,2]

[1]Department of Physics, University of Duisburg-Essen, 47048 Duisburg, Germany.

[2]Center for Nanointegration (CENIDE), 47048 Duisburg, Germany

[3]Fritz-Haber-Institut der Max-Planck-Gesellschaft, Faradayweg 4-6, D-14195 Berlin, Germany

**Abstract:**

In this work, we study the heat transfer from electron to phonon system within a five monolayer thin epitaxial Pb film on Si(111) upon fs-laser excitation. The response of the electron system is determined using time-resolved photoelectron spectroscopy while the lattice excitation is measured by means of the Debye-Waller effect in time-resolved reflection high-energy electron diffraction. The electrons lose their heat within 0.5 ps while the lattice temperature rises slowly in 3.5 to 8 ps, leaving a gap of 3-7 ps. We propose that the hidden energy is transiently stored in high-frequency phonon modes for which diffraction is insensitive and which are excited in 0.5 ps. Within a three-temperature model we use three heat baths, namely electrons, high-frequency and low-frequency phonon modes to simulate the observations. The excitation of low-frequency acoustic phonons, i.e., thermalization of the lattice is facilitated through anharmonic phonon-phonon interaction.

**Introduction:**

The transfer of energy from electrons to the nuclear degrees of freedom facilitates many processes in nature and technology on an atomic level. The chemical reaction in water splitting in chlorophyll, the protein folding in rhodopsin in human eye, or light emission in a diode are famous examples where energy transfer on a quantum level is crucial for the functionality of the desired process. Joules heating originating from the electric current through a resistor serves as another fundamental example for the transfer of energy from weakly excited electrons to the lattice system in solid state matter. The electric resistance still can be described in the framework of the Drude model of conductivity through the finite mean path of electrons between subsequent scattering events giving qualitatively correct results. The fundamental process of energy transfer between the two subsystems (electrons and lattice), however, is not well described.

Here, in contrast to considering a quasi-stationary situation, we explore the pathways of energy flow from the electron system to the lattice system subsequent to an impulsive optical excitation through a combined spectroscopy and diffraction study. From the time domain response of both the excited electrons and the subsequently excited lattice we determined the parameters for the energy flow among the subsystems involved: the electron-phonon-coupling parameter can be accessed directly. We observe a fast transfer of energy from the electron system in only 700 fs while the rise of lattice mean squared displacements occurs only after 3 - 6 ps. And the question is: where has the thermal energy been stored within this tame? The "missing" energy is hidden in high frequency phonons at small vibrational amplitude which are not observed in diffraction. This pronounced non-equilibrium situation among three subsystems lasts for > 10 ps and also violates Boltzmanns equipartition theorem among the phonon system. In a pump-probe setup a fs-IR-laser pulse is used as pump for the electron system. The response of the electron system is followed through time-resolved photo electron spectroscopy (tr-PES) while the response of the lattice is followed by means of the Debye-Waller effect in time-resolved reflection high-energy electron diffraction (tr-RHEED).

**Experimental**

The experiments were performed under ultra-high vacuum (UHV) conditions in two experimental setups by tr-PES [1] and tr-RHEED [2]. In both experiments the samples were prepared following the same recipe: the Si(111) substrates were prepared by degassing and flash-annealing at $T > 1200°C$ to desorb the native oxide. Prior to film growth a Pb-$\beta(\sqrt{3}\times\sqrt{3})$ reconstruction with 1/3 ML coverage was prepared through adsorption of Pb at 600°C [3,4]. The growth of ultrathin and continuous epitaxial Pb films of in total 5, 6, and 7 monolayer (1 $ML_{Pb} = 9.5 \times 10^{14}$ atoms/cm$^2$) thickness were monitored by RHEED intensity oscillations during deposition. In order to suppress island formation, we used the kinetic pathway of growth below 100 K and subsequent annealing at 180 K. In addition, we have grown Pb islands of 5 ML thickness following a recipe reported by Huppalo *et al.* [5]: the deposition temperature $T_{dep} > 150$ K defines the thickness of epitaxial islands which are stabilized by quantum-size effects of the electron system. The thickness of the Pb islands was verified by their electronic quantum well states and their cooling time constants to the Si substrate in the tr-PES [6] and tr-RHEED experiment [7], respectively.

The Pb films were excited through fs-IR-laser pulses at 800 nm for various laser fluences $F$. The absorbed energy density $\Phi_{abs}$ in the Pb film was determined from the maximum temperature $T_{max,el}$ of the electron system and the lattice temperature $T_{max,lat}$ after equilibrium. The sample temperatures were $T_0 = 19$ K and 80 K for the tr- RHEED and the tr- PES study, respectively.

The tr-PES measurements were carried out using an amplified Ti : sapphire laser system operating at 300 kHz repetition rate. Its fundamental infrared (IR) output at $h\nu_1 = 1.5$ eV was used to optically excite the sample. The ultraviolet (UV) probe pulses at $h\nu_1 = 6$ eV were generated by frequency quadrupling of the fundamental $h\nu_1$ by two consecutive $\beta$-bariumborate crystals and subsequent recompression by prism pairs. The IR pump pulses had a pulse duration of 55 fs, whereas the UV probe pulse were slightly longer with 80 fs due to nonlinear effects in the quadrupling process. The cross-correlation of pump and probe pulses resulted in an overall time resolution of 100 fs. Typical absorbed pump fluences were of the order of $F_{abs} = 50-1000$ µJ/cm$^2$ and thus below the damage threshold of Pb/Si(111) since no irreversible spectral changes were encountered [8].

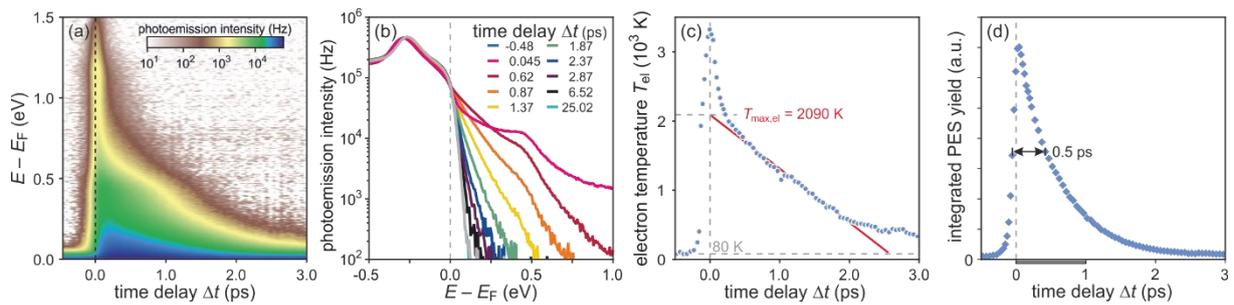

**Figure 1:** tr- PES. (a) false color plot of photoemission density as function of energy above $E_F$ and time delay $\Delta t$. (b) tr-PES spectra from a 5 ML Pb film on Si(111) at 80 K prior and after fs-laser optical excitation at an absorbed energy density of 6.3 µJ/cm$^2$. Different time delays are indicated. (c) From the slope of the photoelectron spectra the evolution of electron temperature $T_{el}$ ($\Delta t$) is determined. Extrapolation of the linear regime to $\Delta t = 0$ gives a maximum electron temperature $T_{max,el} = 2.1 \times 10^3$ K. (d) Electronic heat content $H_{el}$ as function of time delays $\Delta t$. The temporal half width is 0.5 ps.

Figure 1(a) shows the dynamics of the photoemitted electrons in a false color representation of the tr-PES intensity as a function of the pump-probe delay. A sharp peak up to 1.5 eV reflects the initial non-thermal excitation in the electron system. Thermalization sets in on 200 fs timescale.

- 2 -

Photoemission spectra are shown in Fig. 1(b) for various time delays $\Delta t$ ranging from -0.48 ps to 6.52 ps. Upon optical excitation enhanced spectral weight is observed for $E > E_F$. The slope of the spectra becomes steeper with increasing $\Delta t$, i.e., the electron system is cooling. The linear slope of the photoelectron yield in the logarithmic scale reflects the underlying Fermi-Dirac distribution of the excited electron system. The broad feature at $E$ = 0.4 eV is caused by a quantum-well state in the Pb film with slightly increased density of states [9,10,11]. Fitting these features to the spectra we obtain the evolution of electron temperature $T_{el}(\Delta t)$ as shown in Fig. 1(c). The temporal evolution is determined by the initial thermalization among the electrons in less than 200 fs [6]. Such linear decay of electron temperature $T_{el}$ is predicted by 2TM during the absence of hot electron diffusion [12], see supporting Fig. S1. The linear extrapolation to $\Delta t$ = 0 gives $T_{max,el} = 2.1 \pm 0.1 \times 10^3$ K. The temporal half width for electronic cooling is obtained at $\tau_{el,T}(T_{max,el}/2)$ = 1.4 ps.

The temporal evolution of the electronic heat content $H_{el}(t)$ can be determined through integration of the photoelectron yield $\int_{E_F}^{\infty} yield(E - E_F) \cdot (E - E_F)\, dE$ and is shown in Fig.1(d). Initially we observe a parabolic decay as expected from the linear drop of $T_{max,el}$ and the electronic heat capacity. The electronic heat capacity is $c_{el} = R\pi^2 T_{el}/2T_F$, where $R$ = 8.314 J/Kmol and $T_F = 1.1 \times 10^5$ K are the molar gas constant and the Fermi temperature of electrons in bulk Pb, respectively. Thus, heating the electrons of a 5 ML thin Pb film to $T_{max,el} = 2.1 \times 10^3$ K requires an absorbed energy density of 6.4 µJ/cm$^2$. The temporal half width of the electrons heat content is only $\tau_{el,heat}$ = 0.5 ps.

The lattice dynamics following the optical excitation was studied by tr-RHEED in a pump-probe scheme using an amplified Ti : sapphire laser system operating at 5 kHz repetition rate. Its fundamental infrared (IR) output at $h\nu_1$ = 1.55 eV and a pulse length of 100 fs was used to optically excite the sample. Ultrashort electron pulses were generated by linear photo emission from a back- illuminated Au photocathode after frequency tripling of the fundamental IR-pulse. The electrons were accelerated to 30 keV, focused by a magnetic lens, and diffracted at the sample under a grazing incidence of 2.5° to ensure surface sensitivity [2, 13]. A tilted-pulse-front scheme compensates the velocity mismatch between the normal incidence fs-laser pump and the grazing electron probe [14]. The temporal overlap of the pulses at $\Delta t$ = 0 together with the temporal resolution of 1.4 ps FWHM were determined through the response of a symmetry-breaking insulator- to-metal phase transition in the Si(111)-In atomic wire system which occurs in less than 700 fs subsequent to the optical excitation [15]. Typical incident pump fluences were of the order of $F_{inc}$ = 0.2–3 mJ/cm$^2$ and thus below the damage threshold of Pb/Si(111) as no irreversible changes were observed. The dynamics of the lattice was analyzed by means of the Debye-Waller effect $I/I_0 = \exp{-\langle(\mathbf{\Delta u}\cdot\mathbf{k})^2\rangle}$ in diffraction with the change of vibrational displacements of the atoms $\mathbf{\Delta u}$ and the momentum transfer $|\mathbf{k}|$ = 7.6 Å$^{-1}$. Further experimental details are described elsewhere [2, 16].

The diffraction pattern of a 5 ML thick Pb film is shown in Fig. 2(a). The presence of elongated streaks confirm diffraction in reflection geometry from the surface of the Pb film. Figure 2(b) depicts the deep drop of the (00) spot intensity upon optical excitation at $\Delta t$ = 0. Five different absorbed energy densities $\Phi_{abs}$ have been applied. The time constants $\tau_{exc}$ of the exponential drop (with the fit shown as solid lines in Fig. 2(b)) decreases from 8 to 3.5 ps with increasing $\Phi_{abs}$ as clearly seen in Fig. 2(c) for the normalized intensity changes. The slow drop of intensity reflects the increasing mean squared displacements $\langle|\mathbf{\Delta u}|^2\rangle$ of the surface atoms upon impulsive optical excitation of the electron system. For the case of $\Phi_{abs}$ = 6.4 µJ/cm$^2$ we obtain $\langle|\mathbf{\Delta u}|^2\rangle$ = 0.028 Å$^2$, i.e., a maximum rise of lattice temperature $\Delta T_{max}$ = 45K from $T_0$ = 19 K to $T_{max}$ = 64 K is reached [17].

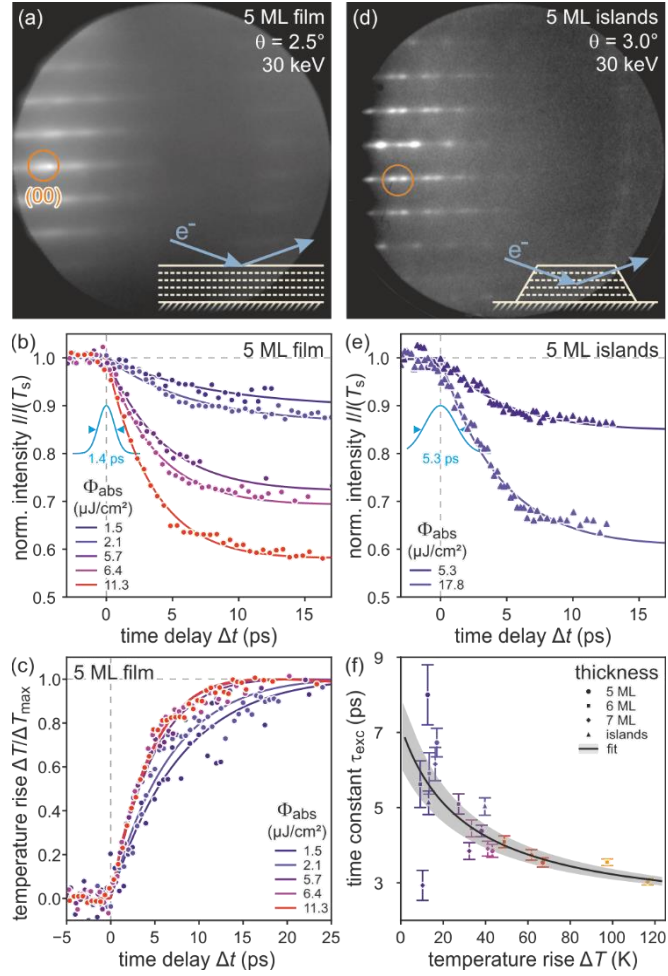

**Figure 2:** (a) RHEED pattern of a 5 ML thick Pb film. Elongated streaks indicate diffraction in reflection from the surface. (b) Transient drop of intensity of the (00) spot for various absorbed energy densities $\Phi_{abs}$. The temporal response function is shown as cyan Gaussian. (c) The normalized temperature rise $\Delta T(\Delta t)/\Delta T_{max}$ clearly shows the change of excitation time constant $\tau_{exc}$ with increasing $\Phi_{abs}$. (d) TED pattern in RHEED of Pb islands. The regular arrangement of spots indicates diffraction in transmission through the islands. (e) Transient drop of intensity of the (00) spot for two different $\Phi_{abs}$ in case of the Pb islands. Solid lines give an exponential fit to the drop of intensity, i.e., a rise of vibrational displacements. (f) Time constants $\tau_{exc}$ of exponential drop of intensity for films of 5, 6, and 7 ML thickness and 5 ML thick islands. The black line gives a guide to the eye to the data. All experiments were performed at a sample temperature $T_0 = 19$ K.

Such long excitation time constants $\tau_{exc}$ were observed also for Pb films of other thickness as shown in Fig. S2 of supplemental material for 6 and 7 ML films. Also, the decay of time constant $\tau_{exc}$ with maximum temperature rise $\Delta T_{max}$, i.e., increasing absorbed energy density $\Phi_{abs}$, is confirmed and summarized in Fig. 2(f) for 5, 6, and 7 ML Pb films.

The same behavior of slow rise of $\langle|\Delta \mathbf{u}|^2\rangle$ is not only observed for thin continuous films but also for thin Pb islands. The regular pattern of sharp diffraction spots in Fig. 2(d) originates from diffraction in transmission. Due to the electrons grazing incidence of 3.0° in this experiment (momentum transfer $|\mathbf{k}| = 9.3$ Å$^{-1}$) with a foreshortening factor of 20 the bulk of the 5 ML thick Pb islands is probed. The intensity drops for two values of $\Phi_{abs}$ are shown in Fig. 2(e) and exhibits the same behavior as for the continuous films. The time constants $\tau_{exc}$ are of the order of 5 ps and are plotted as triangles in Fig. 2(f).

Summarizing, we observe a consistently slow rise of $\langle|\Delta \mathbf{u}|^2\rangle$ on the order of 3 to 7 ps which is independent on the thickness of the Pb films, observed both for surface and bulk, and becomes slightly faster with increasing excitation density.

**Discussion**

Obviously, the time scales for loss of heat from the electrons system in 0.5 ps and rise of temperature in the lattice system in 4 ps do not match at all. This raises the question of where the missing energy is hidden for 3 ps? We can exclude effects through low temporal resolution in diffraction as the FWHM of the temporal response function is 1.4 ps as shown in Fig. S3 of the supporting material. The missing energy can also not be hidden in a spin system for several picoseconds [18] or any kind of latent heat since Pb is nonmagnetic and does not exhibit any structural phase transition, respectively. This leaves the phonon system as the remaining reservoir to hide the energy for more than 3 ps! Here, we propose the initial generation of high frequency phonons $\omega_{hf}$ [19] which subsequently decay into low frequency phonons $\omega_l$ at the $\Gamma$ point.

Because the displacement amplitude $u(\omega)$ of a phonon is inversely proportional to its frequency $\omega$ the Debye-Waller effect becomes insensitive to the population of such zone-boundary phonons with high frequency $\omega_{hf}$ and small amplitude $u(\omega_h)$ compared to zone center phonons with low frequency $\omega_{lf} \rightarrow 0$ and large amplitude $u(\omega_{lf} \rightarrow 0)$[20]. Let assume a given amount of heat $H_{el}$ in the electron system arising from the absorbed energy density $\Phi_{abs}$ = 6.4 µJ/cm$^2$. The heat $H_{el}$ is then redistributed to high frequency phonons with $\omega_{hf}$. The number of high frequency phonons is thus given by $N_{hf} = H_{el}/\hbar\omega_{hf}$. Since phonons are bosons, incoherent superposition results in an MSD $\langle u_{hf}^2\rangle \propto N_{hf}/\omega_{hf}^2 = H_{el}/\hbar\omega_{hf}^3$. These high frequency phonons now decay into low frequency phonons at the zone center with $\omega_{lf}$. Energy conservation gives the number of low frequency phonons $N_{lf} = N_{hf}\cdot\omega_{hf}/\omega_{lf}$. This results in an MSD $\langle u_{lf}^2\rangle \propto N_{lf}/\omega_{lf}^2 \propto N_{hf}\cdot\omega_{hf}/\omega_{lf}^3 = H_{el}/\hbar\omega_{lf}^3$. Assuming the conservation of energy, the MSD varies with the third power of the phonon frequency $\langle u^2\rangle \propto 1/\omega^3$. Under such conditions the Debye-Waller effect is blind to the high frequency phonons [20].

Under equilibrium conditions the temperature is determined via the Debye-Waller effect which is proportional to the mean squared amplitude of the atom's displacements: $I/I_0 = \exp{-1/3\langle u^2\rangle k^2}$. However, under the non-equilibrium conditions resulting from the emission of phonons during cooling of the excited electrons, the equipartition in the phonon phase space is inevitably violated [19, 21] and thus the Debye-Waller effect no longer is a good measure of lattice excitation.

The emission probability of phonons with large wavevector **q** and high frequency $\omega_{hf}$ increases with $|\mathbf{q}|$ since more combinations of $\Delta\mathbf{k} = \mathbf{q}$ and $\Delta E(\Delta\mathbf{k}) = \hbar\omega_{hf}$ across the 3D electronic band structure become possible [22, 23]. In contrast, the emission of low frequency acoustic phonons at the $\Gamma$ point becomes increasingly unlikely with $\Delta\mathbf{k} = \mathbf{q} \rightarrow 0$ since electronic states $E_i(\mathbf{k})$ with $E_1(\mathbf{k}) - E_2(\mathbf{k} - \mathbf{q} \rightarrow 0) = \hbar\omega_{lf}$ become very rare and leaves only intra band transitions as possible mechanism. Those, however, are very improbable, as the slope of the dispersion of acoustic phonons d $\hbar\omega_{lf}$/d $q \cong 10$ meV/Å$^{-1}$ and those of typical electronic states d $E(\mathbf{k})$/d $k \cong 1$ eV/Å$^{-1}$ typically differ by two orders of magnitude or more, rendering this process impossible without breaking conservation of energy and momentum [22].

In order to gain more insight into the flow of energy through the electronic and phonon subsystems we performed simulations using a three-temperature model [21]. The temperatures $T_{el}$, $T_{hf}$, and $T_{lf}$ of the subsystems are modeled by three rate equations:

$$\frac{\mathrm{d}}{\mathrm{d}t}T_{el} = \frac{1}{c_{el}}[A(t) - g_{el-hf}\cdot(T_{el} - T_{hf})], \quad (2a)$$

$$\frac{\mathrm{d}}{\mathrm{d}t}T_{hf} = \frac{1}{c_{l,hf}}[g_{el-hf}\cdot(T_{el} - T_{hf}) - g_{hf-lf}\cdot(T_{hf} - T_{lf})], \quad (2b)$$

$$\frac{\mathrm{d}}{\mathrm{d}t}T_{lf} = \frac{1}{c_{l,lf}}g_{hf-lf}\cdot(T_{hf} - T_{lf}), \quad (2c)$$

with $A(t)$ describing the optical irradiation, the temperature-dependent lattice heat capacities $c_{l,hf}$ and $c_{l,lf}$ for the high and low frequency phonons, respectively, and the coupling parameters $g_{i-j}$ with i,j indicating the electron, high and low frequency phonon system. Due to the missing fast initial drop of the RHEED intensity we conclude that the low frequency phonons are not directly excited by the electrons and safely set $g_{el-lf} = 0$. Thus, from the decay of $T_{el}$ the coupling parameter $g_{el-hf} = 3.3e16$ W/m³K was directly determined. The coupling parameter between the two phonon subsystems $g_{hf-lf} = 12e16$ W/m³K was then determined by fitting the 3TM to the diffraction data shown in Fig. 2(c). It is worth noting here that the differences between heat capacities of electron, ph$_{hf}$ and ph$_{lf}$ subsystems determine how strong the coupling constants for electron- ph$_{hf}$ and ph$_{hf}$-ph$_{lf}$ are. In other words, the differences between heat capacities would act as bottleneck on the energy transfer between these subsystems.

Figure.3 compares the 3TM simulations with the experimental results of heat $H_{el}$ (blue dots) and temperature in the lattice system (green dots and line). The simulated temperature $T_{hf}$ of the high frequency phonon system is depicted as red solid line. The initial rise of the population of high frequency phonons in $\tau_{hf}$ = 0.7 ps reflects the fast decay of heat $H_{el}$. In any case we observe an overshooting of $T_{hf}$ until thermalization is reached for $t$ > 15 ps.

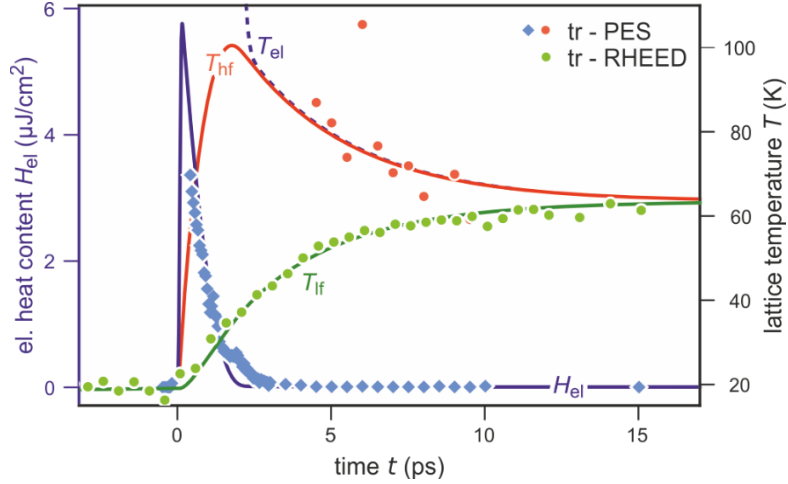

**Figure 3:** 3TM simulations for electronic heat $H_{el}$ (blue) and lattice temperatures for the high frequency (red) and low frequency (green) subsystems as solid lines. tr-PES and tr-RHEED data are shown as solid blue diamonds and green dots, respectively. Red dots depict the transient temperature $T_{el}$ of the electron system serving as a measure for the temperature $T_{hf}$ of the high frequency phonon system.

Due to temperature proportionality of the electron's heat capacity $c_{el} \sim T$ the electron temperature $T_{el}$ (shown as blue dashed line in Fig. 3) merges $T_{hf}$ at ~3 ps and then follows the temperature of the hidden high frequency phonon system since ($T_{el}$ - $T_{hf}$) ~ 0 in Eq. 2(a). Thus, for $t$ > 3 ps the experimentally measured temperature of the electron system $T_{el}$ (shown as red dots in Fig. 3) provides a measure of the temperature $T_{hf}$ of the high frequency phonon system. We corrected the values of $T_{el}$ for the difference in sample temperature of the two experiments. Now, such values for $T_{hf}$ allows us to determine the ratio of the heat capacities $c_{l,hf}/c_{l,lf}$ = 0.4 / 0.6 for the two phonon subsystems with $c_{l,hf}$ + $c_{l,lf}$ = $c_l$ the lattice heat capacity of lead.

The fluence dependence of the excitation time constant as shown in Fig. 2(f) corroborates our proposed paths of energy flow: initially the high frequency phonons are emitted from the hot electron system. Subsequently the acoustic low frequency phonons are populated through anharmonic coupling between high and low frequency phonons. With increasing temperature this process becomes more efficient as the atoms start to sample increasingly anharmonic regimes of the potential energy surface. Thus, we can exclude an emission of low frequency phonons directly from the excited electron system as the electron phonon coupling does not exhibit such a strong temperature dependence [24].

**Conclusion:**

In summary, the heat transfer between electronic and phononic subsystems for a 5 ML Pb film was investigated using tr-PES and tr-RHEED. Such comprehensive investigations showed that the electrons dissipated their heat in less than 0.5 ps, while the lattice heating occurred slowly in 3.5-8 ps. Simulation of such observations using the three-temperature model indicated that the hidden heat within this timescale could transiently be stored in high-frequency phonon modes at the zone boundary where the tr-RHEED was insensitive to detect. Our model rationally explains that the thermalization of the lattice is effectively due to anharmonic phonon-phonon interactions

**Acknowledgment:**

We gratefully acknowledge fruitful discussions with B. Rethfeld, P. Kratzer, and M. Gruner. This work was funded by the Deutsche Forschungsgemeinschaft (DFG, German Research Foundation) Project No. 278162697-SFB 1242.